\documentclass[aps,prb,twocolumn,showpacs,superscriptaddress,amsmath,amssymb]{revtex4}

\usepackage{graphicx}
\usepackage{dcolumn}
\usepackage{bm}
\usepackage{color}

\RequirePackage[russian,english]{babel}

\begin{document}

\title{Oblique surface Josephson plasma waves in layered superconductors}

\author{Yu.O.~Averkov}
\affiliation{ A.Ya.~Usikov Institute for Radiophysics and Electronics, National Academy of Sciences of Ukraine, 61085 Kharkov, Ukraine}

\author{V.M.~Yakovenko}
\affiliation{ A.Ya.~Usikov Institute for Radiophysics and Electronics, National Academy of Sciences of Ukraine, 61085 Kharkov, Ukraine}

\author{V.A.~Yampol'skii}
\affiliation{ A.Ya.~Usikov Institute for Radiophysics and Electronics, National Academy of Sciences of Ukraine, 61085 Kharkov, Ukraine}
\affiliation{V.N.~Karazin Kharkov National University, 61077 Kharkov, Ukraine}
\affiliation{Advanced Science Institute, RIKEN, Saitama, 351-0198, Japan}

\author{Franco Nori}
\affiliation{Advanced Science Institute, RIKEN, Saitama, 351-0198, Japan}
\affiliation{Department of Physics, University of Michigan, Ann Arbor, MI 48109, USA}

\begin{abstract}
We have theoretically studied oblique surface waves (OSWs) which propagate along the interface between a dielectric and a layered superconductor. We assume that this interface is perpendicular to the superconducting layers, and OSWs at the interface can propagate at an arbitrary angle with respect to them. The electromagnetic field of the OSWs in a layered superconductor is a superposition of an ordinary wave (with its \emph{electric} field parallel to the layers) and an extraordinary wave (with its \emph{magnetic} field parallel to the layers). We have derived the dispersion equation for the OSWs and shown that the dispersion curves have end-points where the extraordinary mode transforms from evanescent wave to bulk wave, propagating deep into the superconductor. In addition, we have analytically solved the problem of the resonance excitation of the OSWs by the attenuated-total-reflection method using an additional dielectric prism. Due  to the strong current anisotropy in the boundary of the superconductor, the excitation of the OSWs is accompanied by an additional important phenomenon: the electromagnetic field component with the orthogonal polarization appears in the wave reflected from the bottom of the prism. We show that, for definite optimal combinations of the problem parameters (the wave frequency, the direction of the incident wave vector, the thickness of the gap between dielectric prism and superconductor, etc.), there is a complete suppression of the reflected wave with its polarization coinciding with the polarization of the incident wave. Contrary to the isotropic case, this phenomenon can be observed even in the dissipationless limit. In such a regime, the complete transformation of the incident wave into a reflected wave with orthogonal polarization can be observed.
\end{abstract}

\pacs{74.78.Fk, 74.50.+r, 74.72.-h}




\maketitle

\section{Introduction}

Surface electromagnetic waves represent a specific kind of macroscopic perturbations which propagate along interfaces between different media. The electromagnetic field in such waves decay exponentially away from the interface deep into both media. A very important example of surface waves is the so-called plasmon-polaritons observed at the interfaces between a normal metal and a dielectric at frequencies about the plasma frequency $\omega_p$, which belongs to the optical or far-infrared ranges~\cite{platzman,agr,wood,Petit,zayats}. Plasmon-polaritons are central in numerous resonance phenomena. Examples of these are: (i) the extraordinary transmission of light through metal films with subwavelength holes~\cite{zayats,ebb}; (ii) an ``inverse'' effect of the resonant suppression of light transmission through modulated metal films with thicknesses less than the skin-depth~\cite{katz}; (iii)  the Wood anomalies in the reflectivity~\cite{agr,wood,Petit,katsspevak02} and transmissivity~\cite{wood1,ebb,2,4,33,kn1,kn2,zayats} of periodically-corrugated metal samples. These optical anomalies could have potential applications for photovoltaics, light control, filtering and detection of radiation in far-infrared and visible frequency ranges.

It would be very desirable to similarly control the electromagnetic radiation in the terahertz (THz) frequency range. The mastering of this range (0.3--10 THz) is a rapidly developing area of research due to promising applications. In principle, plasmon-polaritons can exist at $\omega \ll \omega_p$. However, the dispersion curves of plasmon-polaritons are very close to the light-line $\omega = c\kappa/\sqrt{\varepsilon_d}$ in the THz frequency range. In this case, the surface waves are almost extended~\cite{agr}, in the sense that they are very weakly localized in the dielectric over distances of about 1~meter (here $\omega$ and $\kappa$ are the frequency and wave vector of plasmon-polaritons, $\varepsilon_d$ is the permittivity of the dielectric, and $c$ is the speed of light). Therefore, most of the electromagnetic energy flows out of the sample, leading to strong radiation losses. To overcome this disadvantage, new systems, e.g., layered superconductors instead of metals, should be considered in order to observe surface waves.

Layered superconductors are strongly anisotropic media where the current along the  crystallographic \textbf{ab}-plane is similar to the current in bulk superconductors, whereas  transport along the $\mathbf{c}$-axis is caused by the intrinsic Josephson effect through the layers (see, e.g., Refs.~\onlinecite{Kl-Mu,Kl-Mu2}). Therefore, layered superconductors are characterized by two different plasma frequencies~\cite{negref}. One of them is of the order of $\omega_p$ for normal metals, but the other one is much smaller and it coincides with the so-called \emph{Josephson plasma frequency} $\omega_J \sim 1$~THz. As was shown in Refs.~\onlinecite{Thz-rev,surf}, the lower plasma frequency, $\omega_J$, defines the spectrum of surface waves in layered superconductors. Namely, due to this feature the dispersion curves of the surface waves in layered superconductors, in contrast to normal metals, deviate substantially from the light-line, and these waves are well localized in the THz frequency range. Surface Josephson plasma waves (SJPWs) can propagate along the interface between the external dielectric and layered superconductor for both geometries, when the superconducting layers are parallel to the sample surface or are perpendicular to it. As shown in Refs.~\onlinecite{surf,B1,neg-ref} the spectrum of SJPWs propagating along the layers consists of two branches, one above $\omega_J$ and the other below it. The spectrum of surface electromagnetic waves propagating across the layers was predicted and studied analytically in Refs.~\onlinecite{perp1,perp2}.

\begin{figure}
\includegraphics [width=8.0 cm,height=5.7 cm]{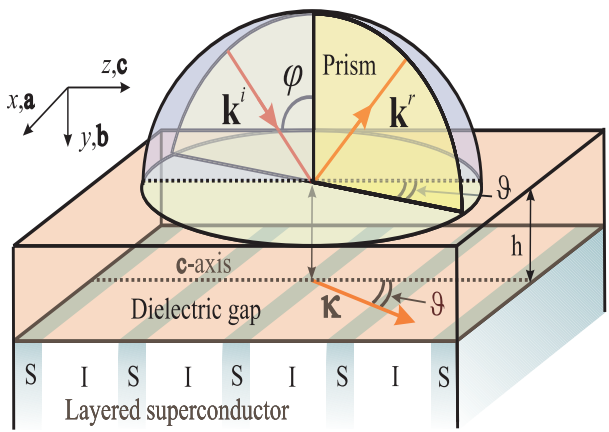}
\caption{\label{Fig1} (Color online) Schematic geometry for the excitation of oblique surface waves. The wave vectors of the incident, reflected, and oblique surface waves are denoted as ${\bf k}^i$, ${\bf k}^r$, and $\bm{\kappa}$, respectively. Note that here S=superconductor and I=insulator.}
\end{figure}

In this paper, we study theoretically the excitation of the oblique surface waves (OSWs) propagating at some angle $\vartheta$ with respect to the crystallographic \textbf{c}-axis, which is assumed to be parallel to the interface between a dielectric and a layered superconductor. We analyze the so-called attenuated-total-reflection (ATR) method for the OSW excitation using an additional dielectric prism with permittivity $\varepsilon_p$. The external electromagnetic wave from a dielectric prism is incident on a superconductor separated from the prism by a thin dielectric gap with permittivity $\varepsilon_d < \varepsilon_p$ (see Fig.~\ref{Fig1}). In the absence of a superconductor, the incident wave completely reflects from the bottom of the prism, if the incident angle $\varphi$ exceeds the limit angle $\varphi_t$ for total internal reflection. However, \emph{the evanescent wave penetrates under the prism} a distance about a wavelength, reaching the layered superconductor and penetrating it. The wave vector of the evanescent mode is along the bottom surface of the prism and its value is higher than $\omega \sqrt{\varepsilon_d} /c$. This feature is the same as for surface waves. Thus, for a certain resonance incident angle, the spatial-and-temporal matching (the frequencies and wave vectors coincide) of the evanescent wave and the surface wave takes place.

It is important to note that the excitation of the oblique surface wave is accompanied by another important resonance phenomenon. Evidently, the anisotropy of the current-carrying capability along the surface of the superconductor results in a conversion of the polarization of the terahertz radiation after reflection from the boundary of the layered superconductor. If the incident wave has, e.g., a transverse magnetic (TM) polarization (a wave with the \emph{magnetic} field parallel to the sample surface), the reflected electromagnetic field contains a transverse electric (TE) wave (with the \emph{electric} field parallel to the sample surface). Moreover, as was shown in Ref.~\onlinecite{Av}, the {\it complete} transformation of the wave polarization can be observed at an appropriate choice of the direction of the incident wave vector.

We stress that this phenomenon is different from the Brewster effect. In the latter, light with TM polarization, for a definite incidence angle, is completely transmitted through a transparent dielectric surface, and, thus, the reflected light has only TE polarization. However, in the phenomenon presented here, the reflected wave is TE-polarized when the incoming wave \textit{does not contain the wave with this polarization}. In other words, the reflected wave is TE-polarized because of the \textit{conversion} of the incident TM wave, but not because of the \textit{separation} of the TE-polarized wave from the \textit{mixed} (TM+TE) incoming wave.

The ratio of the amplitudes of the reflected TE and TM waves is controlled by the wave frequency $\omega$, angles $\vartheta$ and $\varphi$, and the thickness $h$ of the dielectric gap. The main point is the to choose the angles $\vartheta$ and $\varphi$ for the incident wave such that the amplitude of the TM reflected wave vanishes. It is known that, in the  case of an \emph{isotropic} metal, one can select the optimal value of the gap thickness $h$ to provide the complete suppression of the reflected wave in resonance. Similarly, in the case of an \emph{anisotropic} superconductor, we can select the best value of the angle $\vartheta$ to provide the complete suppression of the reflected TM wave in resonance. In the latter case, one part of the incident energy flux dissipates in the layered superconductor and the other part is reflected as the TE wave. The theoretical study of the interplay of the OSWs excitation and the transformation of the wave polarization after its reflection from the boundary of the layered superconductor is the main goal of this paper.

Thus, a layered superconductor plays two important roles in the phenomenon studied in this paper, which a simple metal or an isotropic superconductor cannot. First, due to the smallness of the Josephson current across the layers, the excited oblique surface waves are well localized in the terahertz frequency range. Second, due to the strong current anisotropy in the surface of the layered superconductor, the excitation of oblique surface waves is accompanied by the transformation of the incident wave polarization after its reflection. Moreover, the complete transformation of the polarization can be achieved by the appropriate choice of the direction of the wave incidence.

In the next sections, we derive and analyze the dispersion relation for the oblique surface waves, and then study analytically the OSWs excitation by the ATR method and the transformation of the electromagnetic wave polarization. For this purpose, we calculate the electromagnetic field in the dielectric prism, in the dielectric gap, and in the layered superconductor, by solving the Maxwell equations with appropriate material equations and boundary conditions. The electromagnetic field in the dielectric prism is presented as a sum of the incident TM-polarized wave and two reflected waves with TM and TE polarizations. In the dielectric gap, we consider four evanescent waves, namely, increasing and decreasing TM- and TE-polarized waves. The evanescent field in the layered superconductor consists of the ordinary and extraordinary decreasing waves. Then, using the continuity conditions for the tangential components of the electric and magnetic fields at the two interfaces (i.e., the dielectric gap-dielectric prism and the layered superconductor-dielectric gap), we express all wave amplitudes via the amplitude of the incident wave. Using these expressions, we derive the reflection coefficient $R_{\rm TM}$ for the TM wave and the conversion coefficient $T_{{\rm TM} \rightarrow {\rm TE}}$ for the mode transformation from TM to TE, as well as the expression for the portion $A$ of the energy flux which comes to the excitation of the oblique surface wave and then dissipates in the superconductor.

\section{Dispersion relation for the oblique surface waves}

\begin{figure}
\includegraphics [width=8.0 cm,height=3.0 cm]{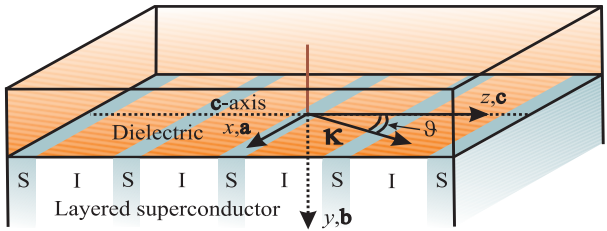}
\caption{\label{Fig2} (Color online) Geometry for the oblique surface waves.}
\end{figure}

Let the interface dividing an isotropic dielectric and a layered superconductor be located in the $y=0$ plane (see Fig.~\ref{Fig2}). We assume that the dielectric region ($y<0$) and the layered superconductor region ($y>0$) are semi-infinite along the $y$-axis. Both media are nonmagnetic ones.

The coordinate system is chosen  in such a way that the $z$-axis is directed along the crystallographic $\mathbf{c}$-axis of the superconductor. We consider oblique surface waves which propagate along the interface between the dielectric and layered superconductor at an angle $\vartheta$ with respect to the \textbf{c}-axis. The electric $\mathbf{E}$ and magnetic $\mathbf{H}$ fields of these waves are presented in the form,
\begin{equation}\label{1}
\mathbf{E}(x,y,z,t),\:\mathbf{H}(x,y,z,t)\propto \exp[i(\bm{\kappa}\bm{\rho} -\omega t)].
\end{equation}
Here $\bm{\rho}=(x,z)$ is the radius vector in the $xz$-plane and $\bm{\kappa}=(k_x, k_z)$ is the wave vector of the oblique surface wave, $|\bm{\kappa}|>k=\omega/c$.

The electromagnetic field in the dielectric is a sum of two terms that correspond to the ordinary and extraordinary evanescent waves,
\begin{equation}\label{e2}
\mathbf{E}(y)=\mathbf{E}^{\rm (o)}(y)+\mathbf{E}^{\rm (e)}(y), \ \:\mathbf{H}(y)=\mathbf{H}^{\rm (o)}(y)+\mathbf{H}^{\rm (e)}(y).
\end{equation}
The magnetic field of the extraordinary wave and the electric field of the ordinary wave are parallel to the crystallographic \textbf{ab}-plane:
\begin{equation}\label{pol}
E^{\rm (o)}_z(y)=H^{\rm (e)}_z(y)=0.
\end{equation}
Using the Maxwell equations, we obtain the relations between the components of the magnetic $\mathbf{H}_d$  and electric $\mathbf{E}_d$ fields for each of these waves:
\[
E^{(\mathrm{o})}_{dy}=-\frac{k_x}{k_{dy}}E^{(\mathrm{o})}_{dx},\quad H^{(\mathrm{o})}_{dx}=\frac{ k_x k_z}{k k_{dy}}E^{(\mathrm{o})}_{dx},
\]
\begin{equation}\label{HTS2d}
H^{(\mathrm{o})}_{dy}=\frac{k_z}{k}E^{(\mathrm{o})}_{dx},\quad
H^{(\mathrm{o})}_{dz}=-\frac{k_x^2+k_{dy}^2}{k k_{dy}}E^{(\mathrm{o})}_{dx},
\end{equation}
\[
E^{(\mathrm{e})}_{dy}=\frac{k_{dy}}{k_x}E^{(\mathrm{e})}_{dx},\quad E^{(\mathrm{e})}_{dz}=-\frac{k_x^2+k_{dy}^2}{k_x k_z}E^{(\mathrm{e})}_{dx},
\]
\begin{equation}\label{HTS11ad}
H^{(\mathrm{e})}_{dx}=-\frac{k_{dy}}{k k_x k_z}\varepsilon_d E^{(\mathrm{e})}_{dx},\quad H^{(\mathrm{e})}_{dy}=\frac{k}{k_z}\varepsilon_d E^{(\mathrm{e})}_{dx},
\end{equation}
Here $k=\omega /c$. The ordinary and extraordinary waves decay deep into the superconductor according to the same exponential law,
\begin{equation}\label{e3}
\mathbf{E}_d^{\rm (o)}(y), \;\mathbf{E}_d^{\rm (e)}(y), \;\mathbf{H}_d^{\rm (o)}(y), \;\mathbf{H}_d^{\rm (e)}(y) \propto \exp(ik_{y}^d y),
\end{equation}
with imaginary normal component of the wave vector,
\begin{equation}\label{k-vac}
k_{dy}=-i(k_x^2 + k_z^2 - \varepsilon_d k^2)^{1/2}.
\end{equation}

The electromagnetic field in the layered superconductor is related to the spatial distribution of the gauge-invariant phase difference $\chi$ of the order parameter between neighboring layers. This phase difference can be found by a solution of a set of coupled sine-Gordon equations (see, e.g., Refs.~\onlinecite{Thz-rev,sak,bul,koy,art,kim,rev2}). In the continuum limit, $\chi$ can be excluded from the set of equations for the electromagnetic fields. In this case, the layered superconductors can be described in terms of usual macroscopic electrodynamics with an anisotropic frequency-dependent dielectric permittivity with components $\varepsilon_c$ and $\varepsilon_{ab}$ across and along the layers, respectively. Indeed, according to Ref.~\onlinecite{negref}, the solution of the set of coupled sine-Gordon equations in the continuum limit is equivalent to the solution of the Maxwell equations for the uniform anisotropic medium with
\[
\varepsilon_{c}(\Omega)=\varepsilon_s\left(1-\frac{1}{\Omega^2}+i\frac{\nu_{c}}{\Omega}\right),
\]
\begin{equation}\label{epsilon2}
\varepsilon_{ab}(\Omega)=\varepsilon_s\left(1-\gamma^2\frac{1}{\Omega^2}+i\gamma^2\frac{\nu_{ab}}{\Omega}\right).
\end{equation}
Here we use the dimensionless parameters, $\Omega=\omega/\omega_J$, $\nu_{ab}= 4\pi \sigma_{ab}/\varepsilon_s\omega_J\gamma^2$, and $\nu_{c}= 4\pi \sigma_{c}/\varepsilon_s\omega_J$; $\gamma=\lambda_c/\lambda_{ab} \gg 1$ is the current-anisotropy parameter, $\lambda_c=c/\omega_J\varepsilon_s^{1/2}$ and $\lambda_{ab}$ are the magnetic-field penetration depths along and across the layers, respectively. The relaxation frequencies $\nu_{ab}$ and $\nu_{c}$ are defined by the quasi-particle conductivities $\sigma_{ab}$ (along the layers) and $\sigma_{c}$ (across the layers), respectively; $\omega_J = (8\pi e D j_c/\hbar\varepsilon)^{1/2}$ is the Josephson plasma frequency. This characteristic frequency is determined by the maximum Josephson current density $j_c$,  the spatial period $D$ of the layered structure, and the interlayer dielectric constant $\varepsilon_s$.

For the layered superconductors, the Maxwell equations give different laws for the decays of the ordinary and extraordinary waves along the $y$-axis,
\[
\mathbf{E}_{\rm s}^{\rm (o)}(y), \;\mathbf{H}_{\rm s}^{\rm (o)}(y) \propto \exp(ik_{{\rm s}y}^{\rm (o)}y)\:,
\]
\begin{equation}\label{e4}
\mathbf{E}_{\rm s}^{\rm (e)}(y), \;\mathbf{H}_{\rm s}^{\rm (e)}(y) \propto \exp(ik_{{\rm s}y}^{\rm (e)}y)
\end{equation}
with \begin{equation}\label{HTS5}
k_{{\rm s}y}^{\rm (o)}=i(k_x^2+k_z^2-k^2\varepsilon_{ab})^{1/2}\:,
\end{equation}
\begin{equation}\label{HTS5a}
k_{{\rm s}y}^{\rm (e)}=i(k_x^2+k_z^2 \varepsilon_c/\varepsilon_{ab}-k^2\varepsilon_{c})^{1/2}\:.
\end{equation}

The relationships between the field components in the layered superconductor are as follows:
\[
E^{(\mathrm{o})}_{sy}=-\frac{k_x}{k^{(\mathrm{o})}_{sy}}E^{(\mathrm{o})}_{sx},\quad  H^{(\mathrm{o})}_{sx}=\frac{k_x k_z}{k k^{(\mathrm{o})}_{sy}}E^{(\mathrm{o})}_{sx},\
\]
\begin{equation}\label{HTS3}
H^{(\mathrm{o})}_{sy}=\frac{k_z}{k}E^{(\mathrm{o})}_{sx},\quad H^{(\mathrm{o})}_{sz}=-\frac{k_x^2+\bigl(k^{(\mathrm{o})}_{sy}\bigr)^2}{k k^{(o)}_{sy}}E^{(\mathrm{o})}_{sx}
\end{equation}
for the ordinary wave and
\[
E^{(\mathrm{e})}_{sy}=\frac{k_{sy}^{(\mathrm{e})}}{k_x}E^{(\mathrm{e})}_{sx},\quad E^{(\mathrm{e})}_{sz}=-\frac{\varepsilon_{ab}}{\varepsilon_c}\frac{k_x^2+\bigl(k_{sy}^{(\mathrm{e})}\bigr)^2}{k_x k_z}E^{(\mathrm{e})}_{sx},
\]
\begin{equation}\label{HTS222a}
H^{(\mathrm{e})}_{sx}=-\frac{k k_{sy}^{(\mathrm{e})}}{k_x k_z}\varepsilon_{ab}E^{(\mathrm{e})}_{sx},\quad H^{(\mathrm{e})}_{sy}=\frac{k}{ k_z}\varepsilon_{ab}E^{(\mathrm{e})}_{sx}
\end{equation}
for the extraordinary waves.

Using the continuity conditions for the tangential components of the electric and magnetic fields at the interface $y=0$, we obtain the dispersion equation for the oblique surface waves,
\begin{equation}\label{de}
k_z^2 k_{sy}^{(\mathrm{o})} (\varepsilon_{ab}-\varepsilon_d) + \varepsilon_{ab} (k_{dy}-k_{sy}^{(\mathrm{e})})(k_{dy}k_{sy}^{(\mathrm{o})}-k_x^2) = 0.
\end{equation}

\begin{figure}
\includegraphics [width=8 cm,height=6.25 cm]{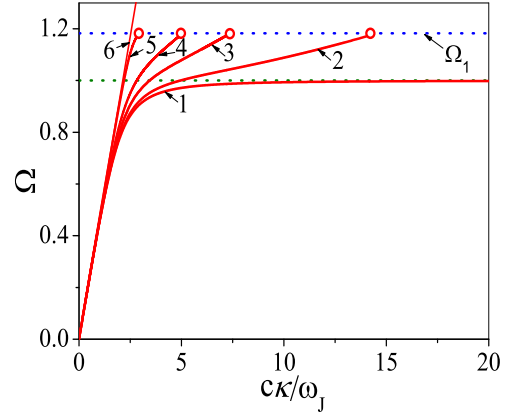}
\caption{\label{Fig3} (Color online) Dispersion curves $\Omega(c\kappa/\omega_J)$ of the OSWs obtained for $\gamma=200$, $\nu_{ab}=\nu_c=0$, $\varepsilon_d=4.56$, and $\varepsilon_s=16$. Curves 1--6 are found for $\vartheta=0^{\circ}$, $10^{\circ}$, $20^{\circ}$, $30^{\circ}$, $60^{\circ}$, and $90^{\circ}$, respectively. The circles at $\Omega = \Omega_{\rm end} \approx\Omega_1=\sqrt{\varepsilon_s/(\varepsilon_s-\varepsilon_d)}$ on the dispersion curves show the end-points. The extraordinary wave is transformed into a bulk wave in these points. When $\kappa\rightarrow\infty$, curve 6 asymptotically approaches the frequency $\Omega=\gamma\sqrt{\varepsilon_s/(\varepsilon_s+\varepsilon_d)}\gg 1$.}
\end{figure}

Figure~\ref{Fig3} shows the numerically calculated dispersion curves of the OSWs for different angles $\vartheta$.
The hollow circles at $\Omega=\Omega_{\rm end} \approx\Omega_1= \sqrt{\varepsilon_s/(\varepsilon_s-\varepsilon_d)}$ show the end-points of the corresponding dispersion curves. In these points, the extraordinary wave, which is present in the oblique surface mode, is transformed into a bulk wave. Indeed, the wave vector $k_{sy}^{(\mathrm{e})}$ vanishes at the end-points and is real and positive for $\Omega>\Omega_{\rm end}$. The analysis of Eq.~(\ref{de}) shows that $\Omega_{\rm end} \approx \Omega_1$ for angles $\vartheta$  not very close to zero or $\pi/2$. For very small $\vartheta \ll 1/\gamma \ll 1$, we have $\Omega_{\rm end} \approx 1$, and $\Omega_{\rm end} \rightarrow\gamma\sqrt{\varepsilon_s/(\varepsilon_s+\varepsilon_d)}\gg 1$ for $(\pi/2-\vartheta) \ll 1/\gamma$.

In the main approximation with respect to the parameter $\gamma \gg 1$ and neglecting the dissipation, Eq.~(\ref{de}) gives
\begin{equation}\label{de-m}
D_0=\frac{1}{k^2}\left(|k_{dy}|^2 +|k_{dy}||k_{sy}^{(\mathrm{e})}| - k_z^2  \right) = 0.
\end{equation}

\section{Excitation of the oblique surface waves using the ATR method}

In this section, we study analytically the OSWs excitation using the ATR method. Let the structure under study be comprised of a dielectric prism with permittivity $\varepsilon_p$, a dielectric gap with thickness $h$ and permittivity $\varepsilon_d$, and a layered superconductor described by a frequency-dependent diagonal tensor of dielectric permittivity with components $\varepsilon_{xx}=\varepsilon_{yy}=\varepsilon_{ab}$ and $\varepsilon_{zz}=\varepsilon_{c}$ given by Eq.~(\ref{epsilon2}) (see Fig.~\ref{Fig1}). All media are supposed to be nonmagnetic. We define the coordinate system so that the prism region occupies the half-space $y<0$, the superconductor region occupies the half-space $y>h$, and the $z$-axis coincides with the crystallographic $\mathbf{c}$-axis of the layered superconductor.

Consider now a TM polarized wave incident from a dielectric prism on a layered superconductor. The incident angle $\varphi$ is larger than the limit angle $\varphi_t =\arcsin[(\varepsilon_d/\varepsilon_p)^{1/2})]$ for the total internal reflection. In this case the incident wave produces a surface wave if the resonance condition is satisfied. This condition implies that the tangential component $k \varepsilon_p^{1/2}\sin \varphi$ of the incident wave vector is equal to the wave vector $\kappa$ of the oblique surface wave,
\[
k\varepsilon_p^{1/2} \sin\varphi=\kappa .
\]

In the prism, the electromagnetic field can be presented as a sum of three terms that correspond to the incident TM polarized wave and two reflected waves with the TE and TM polarizations. So, the $x$-component of the electric field in the dielectric prism has a form,
\[
E_{px}(x,y,z,t) = \left[E^i_x\exp(ik_{py} y) \right.
\]
\begin{equation}\label{eq1}
\left.+ (E^r_{x \:{\rm TM}}+E^r_{x \:{\rm TE}})\exp(-ik_{py} y) \right]\exp\bigl[i(\bm{\kappa}\bm{\rho} -\omega t)\bigr].
\end{equation}
Here $E^i_x$, $E^r_{x \:{\rm TM}}$, and $E^r_{x \:{\rm TE}}$ are the amplitudes of the $x$-components of the electric field in the incident wave and in the reflected TM and TE waves; $k_{py} =k \varepsilon_p^{1/2} \cos\varphi$ is the normal component of the wave vector of the incident wave.

The electric and magnetic field components in the prism can be expressed via $E^i_x$, $E^r_{x \:{\rm TM}}$, and $E^r_{x \:{\rm TE}}$ using the Maxwell equations,
\begin{equation}\label{Test2}
    E_{py}=-\frac{\kappa^2}{k_{py}k_x}\Bigl[E_{x}^i\exp(ik_{py} y)-E^r_{x \:{\rm TM}}\exp(-ik_{py}y)\Bigr],\
\end{equation}
\[
E_{pz}=\frac{k_z}{k_x}E^i_x \exp(ik_{py}y)
\]
\begin{equation}\label{Test3}
     +\Bigl(\frac{k_z}{k_x}E^r_{x \:{\rm TM}}-
    \frac{k_x}{k_z}E^r_{x \:{\rm TE}}\Bigr)\exp(-ik_{py}y),\
\end{equation}
\[
H_{px}=\frac{\varepsilon_p\, k k_z}{k_{x} k_{py}} E^i_x \exp(ik_{py} y)
\]
\begin{equation}\label{Test4}
    +\Bigl(-\frac{\varepsilon_p \,k k_z}{k_{x} k_{py}} E^r_{x \:{\rm TM}}+
    \frac{k_{x} k_{py}}{k k_{z}}E^r_{x \:{\rm TE}}\Bigr) \exp(-ik_{py}y),
 \end{equation}
\begin{equation}\label{Test5}
    H_{py}=\frac{\kappa^2}{k k_{z}}E^r_{x \:{\rm TE}}\exp(-ik_{py}y),
\end{equation}
\[
 H_{pz}=-\frac{k \varepsilon_p}{k_{py}}E^i_x \exp(ik_{py} y)
\]
\begin{equation}\label{Test6}
 +\left(\frac{k\varepsilon_p}{k_{py}} E^r_{x \:{\rm TM}}+\frac{k_{py}}{k}E^r_{x \:{\rm TE}}\right) \exp(-ik_{py}y).
\end{equation}
Here and henceforward, we omit the multiplier $\exp[i(\bm{\kappa}\bm{\rho} -\omega t)]$ in all expressions for the electromagnetic field components.

For the dielectric gap region, we define the electromagnetic field in the same manner as for the prism region, namely, as the superposition of the TM- and TE-polarized waves with the following
components:
\begin{equation}\label{Neq6}
    E_{dx \:{\rm TM}}=B_{1}\exp(i k_{dy} y)+B_2\exp(-i k_{dy} y),
\end{equation}
\begin{equation}\label{Neq7}
    E_{dy \:{\rm TM}}=-\frac{\kappa^2}{k_{dy} k_x}\Bigl[B_1\exp(i k_{dy}y)-B_2\exp(-i k_{dy}y)\Bigr],
\end{equation}
\begin{equation}\label{Neq7Add}
    E_{dz \:{\rm TM}}=\frac{k_z}{k_x}E_{dx \:{\rm TM}},
\end{equation}
\begin{equation}\label{Neq8}
    H_{dx \:{\rm TM}}=-\frac{\varepsilon_d k k_z}{\kappa^2}E_{dy \:{\rm TM}},\
  H_{dz \:{\rm TM}}=\frac{\varepsilon_d k k_x}{\kappa^2} E_{dy \:{\rm TM}},
\end{equation}
\begin{equation}\label{Neq6H}
    E_{dx \:{\rm TE}}=C_{1}\exp(i k_{dy} y)+C_2\exp(-i k_{dy} y),
\end{equation}
\begin{equation}\label{Neq7AddH}
    E_{dz \:{\rm TE}}=-\frac{k_x}{k_z}E_{dx \:{\rm TE}},
\end{equation}
\begin{equation}\label{Neq8H}
    H_{dx \:{\rm TE}}=-\frac{k_x k_{dy}}{k k_z}
    \Bigl[C_1\exp(i k_{dy}y)-C_2\exp(-i k_{dy}y)\Bigr],\
 \end{equation}
\begin{equation}\label{Neq9H}
H_{dy \:{\rm TE}}=\frac{\kappa^2}{k k_z} H_{dx \:{\rm TE}},\
H_{dz \:{\rm TE}}=\frac{k_z}{k_x}H_{dx \:{\rm TE}}.
\end{equation}

The electromagnetic field in the layered superconductor is the superposition of ordinary and extraordinary waves, and it is described by the same formulaes as in the previous section [see Eqs.~(\ref{e4})--(\ref{HTS222a})].

Using the continuity boundary conditions for the tangential components of the electric and magnetic fields at the two interfaces (i.e., the dielectric prism-dielectric gap and the dielectric gap-layered superconductor), we derive eight linear algebraic equations for eight unknown amplitudes (for 4 waves in the dielectric gap, 2 waves in the layered superconductor, and amplitudes $E^r_{x \:{\rm TM}}$, $E^r_{x \:{\rm TE}}$ of the reflected waves in the dielectric prism). Solving these equations, we obtain the reflectivity coefficient
\begin{equation}\label{R}
R_{\rm TM}(\varphi, \vartheta)=\frac{|\textbf{E}^r_{\:{\rm TM}}(\varphi, \vartheta)|^2}{|\textbf{E}^i|^2}
\end{equation}
for the TM wave, the coefficient
\begin{equation}\label{T}
T_{{\rm TM}\rightarrow {\rm TE}}(\varphi, \vartheta)=\frac{|\textbf{E}^r_{\:{\rm TE}}(\varphi, \vartheta)|^2}{|\textbf{E}^i|^2}
\end{equation}
of the transformation of the incident TM wave to the reflected TE wave, and the absorptivity
\begin{equation}\label{A}
A(\varphi, \vartheta)=1-R_{\rm TM}(\varphi, \vartheta)-T_{{\rm TM}\rightarrow {\rm TE}}(\varphi, \vartheta).
\end{equation}
It is suitable to present these coefficients in the form,
\begin{equation}\label{RR}
R_{\rm TM}=\frac{(D_0+D_c)^2+(L_{\rm TM}-L_{\rm TE}-\Gamma)^2}{(D_0+D_c)^2+(L_{\rm TM}+L_{\rm TE}+\Gamma)^2},
\end{equation}
\begin{equation}\label{TT}
T_{{\rm TM}\rightarrow {\rm TE}}=\frac{4L_{\rm TM}L_{\rm TM}}{(D_0+D_c)^2+(L_{\rm TM}+L_{\rm TE}+\Gamma)^2},
\end{equation}
\begin{equation}\label{AA}
A=\frac{4\Gamma L_{\rm TM}}{(D_0+D_c)^2+(L_{\rm TM}+L_{\rm TE}+\Gamma)^2}.
\end{equation}

The expressions for the parameters $D_c$, $L_{\rm TM}$, $L_{\rm TE}$, and $\Gamma$ are very cumbersome. However, they can be significantly simplified for the case of small dissipation and weak coupling of the electromagnetic fields in the layered superconductor and the dielectric prism, when the following inequalities are satisfied:
\begin{equation}\label{Ch}
\Gamma \ll 1, \qquad C_h=\exp(-2|k_{dy}| h) \ll 1.
\end{equation}
For this case, using the smallness of the parameter $\gamma^{-1}$, we obtained the following formulaes:
\[
D_c=2\,C_h \frac{\varepsilon_d ^2\cos ^2\vartheta}{\varepsilon_p-\varepsilon_d}
\]
\begin{equation}\label{DC}
\times \frac{|k_{dy}|^2 k_{py}^2 + \frac{\varepsilon_p}{\varepsilon_d}k^2|k_{dy}|^2 + \frac{\kappa^4 k_{py}^2}{\varepsilon_d k^2} \left(1-\frac{|k_{dy}|^2}{k_z^2} \right)}{\kappa^4 +  |k_{dy}|^2 k_{py}^2}\,,
\end{equation}
\begin{equation}\label{LTM}
L_{\rm TM}=4\,C_h \frac{\varepsilon_d \cos ^2\vartheta}{\varepsilon_p-\varepsilon_d}\frac{\varepsilon_p \, \varepsilon_d \,k^2 |k_{dy}| k_{py}}{\kappa^4 +  |k_{dy}|^2 k_{py}^2}\,,
\end{equation}
\[
L_{\rm TE}=2\,C_h \frac{\varepsilon_d \cos ^2\vartheta}{\varepsilon_p-\varepsilon_d} \frac{k_x^2}{k_z^2}\frac{k^2|k_{dy}|k_{py}}{\kappa^4 +  |k_{dy}|^2 k_{py}^2}
\]
\begin{equation}\label{LTE}
\times \left [\frac{\varepsilon_p}{\varepsilon_d}\frac{\kappa^4}{k^4} +2 (\varepsilon_p+\varepsilon_d)\frac{\kappa^2}{k^2} - \varepsilon_p \, \varepsilon_d \right]\,,
\end{equation}
\[
\Gamma=\Gamma_c + \Gamma_{ab},
\]
\begin{equation}\label{GAB}
\Gamma_c=\nu_c \frac{\varepsilon_s}{2 \Omega}\frac{|k_{dy}|}{|k_{sy}^{(\mathrm{e})}|}\,,
\quad \Gamma_{ab}=\nu_{ab} \frac{\Omega ^2}{2\gamma}\frac{k_x^2(|k_{dy}|+|k_{sy}^{(\mathrm{e})}|)}{k^3}\,.
\end{equation}

\section{Resonance suppression of the reflectivity}

Equation (\ref{RR}) together with Eqs.~(\ref{de-m}), (\ref{DC})--(\ref{GAB}) describes the resonance behavior of the reflectivity coefficient $R_{\rm TM}(\varphi, \vartheta)$ when changing the direction of the incident wave propagation. The resonance suppression of $R_{\rm TM}(\varphi, \vartheta)$ occurs when the tangential component $\kappa$ of the incident wave vector coincides with the wave vector of the oblique surface wave. Indeed, the equation
\begin{equation}\label{D=0}
D_0+ D_c=0
\end{equation}
is the dispersion relation for the OSWs in the system under consideration, which consists of the layered superconductor, the dielectric gap, and the dielectric prism. Because of the coupling of the electromagnetic fields in the superconductor and the dielectric prism, this relation differs slightly from the dispersion equation $D_0=0$ for the OSWs in the system without the prism.

The parameters $L_{\rm TM}$, $L_{\rm TE}$, and $\Gamma$ in the denominator in Eq.~(\ref{R}) define the resonance width. These are related to three channels for the losses of the electromagnetic energy of the OSWs. The term $\Gamma$ describes the dissipation, the term $L_{\rm TM}$ is related to the energy leakage through the dielectric prism in the form of a TM bulk wave, and  the term $L_{\rm TE}$ describes the leakage in the form of a TE wave.

\begin{figure}
\includegraphics [width=8 cm,height=6.5 cm]{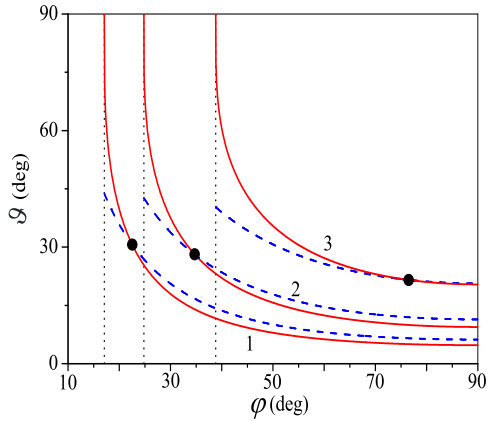}
\caption{\label{Fig4} (Color online) Graphical solutions of the set of equations (\ref{PT1}) and (\ref{PT2}). The red solid curves correspond to the dispersion relation for OSWs and the blue dashed curves are for the condition $L_{\rm TM}=L_{\rm TE}$. The curves start from the angles $\varphi_t$, which are the critical angles for the total internal reflection from the bottom of the dielectric prism. The  material parameters are chosen as $\varepsilon_p=11.6$ (this value corresponds to silicon, which is frequently used as the material for the dielectric prism for the ATR experiments in the THz frequency range~\cite{agr}), $\varepsilon_s = 16$ (this interlayer  permittivity corresponds to $\rm Bi_2Sr_2CaCu_2O_{8+\delta}$~\cite{Thz-rev}). Curves are plotted for $\varepsilon_d=1$ (the vacuum gap, curves 1), $\varepsilon_d=2.04$ (teflon, curves 2), and $\varepsilon_d=4.56$ (quartz, curves 3). Other parameters are $\Omega=1$, $\nu_{ab}=\nu_c = 0$. The intersections of red and blue curves occur at $\varphi= \varphi_{\rm int} \approx 22.5^\circ$, $\vartheta=\vartheta_{\rm int} \approx 30.6^\circ$ (curves labeled by 1); $\varphi_{\rm int} \approx 34.9^\circ$, $\vartheta_{\rm int} \approx 28^\circ$ (curves 2); $\varphi_{\rm int} \approx 76.4^\circ$, $\vartheta_{\rm int} \approx 21.6^\circ$ (curves 3).}
\end{figure}

\begin{figure}
\includegraphics [width=10 cm,height=13.4 cm]{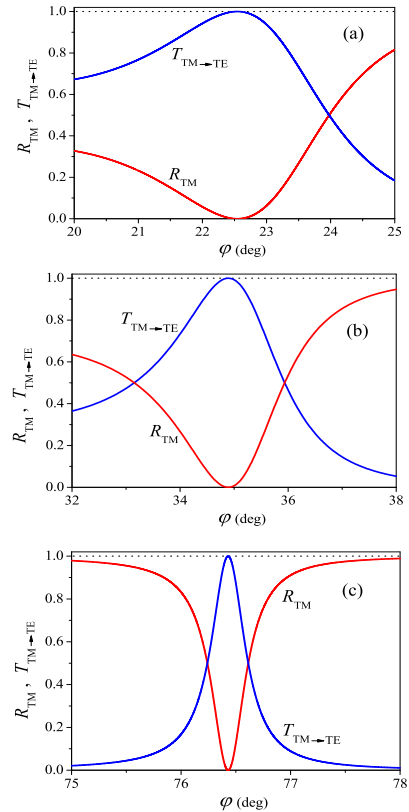}
\caption{\label{Fig5} (Color online) Complete resonance suppression of the reflectivity in the dissipationless regime. The dependence of the reflectivity $R_{\rm TM}$ and the transformation coefficient $T_{{\rm TM}\rightarrow {\rm TE}}$ on the incident angle $\varphi$ are shown by the solid red and dashed blue curves for the same parameters as in Fig.~\ref{Fig4}: (a) $\varepsilon_d=1$, $\vartheta= 30.6^\circ$; (b) $\varepsilon_d=2.04$, $\vartheta= 28^\circ$; (c) $\varepsilon_d=4.56$, $\vartheta= 21.6^\circ$. These three cases correspond to the three intersections in Fig.~\ref{Fig4}. The gap thickness $h$ is equal to $1.2 c/\omega_J$. }
\end{figure}

The resonance suppression of $R_{\rm TM}(\varphi, \vartheta)$ can be complete if both terms in the numerator in Eq.~(\ref{RR}) vanish at the same angles $\varphi$ and $\vartheta$. Namely, the complete suppression of $R_{\rm TM}(\varphi, \vartheta)$ occurs if the dispersion relation Eq.~(\ref{D=0}) and the condition
\begin{equation}\label{CS}
L_{\rm TM}=L_{\rm TE}+\Gamma.
\end{equation}
are fulfilled simultaneously. Recall that, in the isotropic case, the condition for complete resonance suppression of $R_{\rm TM}(\varphi, \vartheta)$ can be realized for metals or superconductors with finite dissipation parameter $\Gamma$ (see, e.g., Ref.~\onlinecite{agr}). However, in the anisotropic case considered here, this phenomenon can be observed even when $\Gamma=0$. In the main approximation with respect to the small parameters $\gamma ^{-1} \ll 1$ and $C_h \ll 1$, the corresponding angles $\varphi$ and $\vartheta$ can be found by solving the set of equations which follow from the equalities $D_0=0$ and $L_{\rm TM}=L_{\rm TE}$,
\[
\sin^2\varphi\cos^2\varphi=\sin^2\varphi-\frac{\varepsilon_d}{\varepsilon_p}
\]
\begin{equation}\label{PT1}
+\sqrt{\sin^2\varphi-\frac{\varepsilon_d}{\varepsilon_p}}
\sqrt{\sin^2\varphi\sin^2\varphi-\frac{\varepsilon_s}{\varepsilon_p}\frac{\Omega^2-1}{\Omega^2}}\,,
\end{equation}
\begin{equation}\label{PT2}
2\cot^2\vartheta=\frac{\varepsilon_p}{\varepsilon_d}\,\frac{\varepsilon_p + \varepsilon_d}{\varepsilon_d}\sin^4\varphi+ \,\frac{2\,\varepsilon_p + \varepsilon_d}{\varepsilon_d}\sin^2\varphi - 1.
\end{equation}
The graphical solutions of these equations are presented in Fig.~\ref{Fig4} for different material parameters.
 Figure~\ref{Fig5} demonstrates the complete resonance suppression of the reflection coefficient that occurs for the dissipationless case, namely at the angles $\varphi= \varphi_{\rm int}$ and $\vartheta=\vartheta_{\rm int}$, which are the solutions of the set of equations (\ref{PT1}), (\ref{PT2}) shown in Fig.~\ref{Fig4}.

\begin{figure}
\includegraphics [width=10 cm,height=16.7 cm]{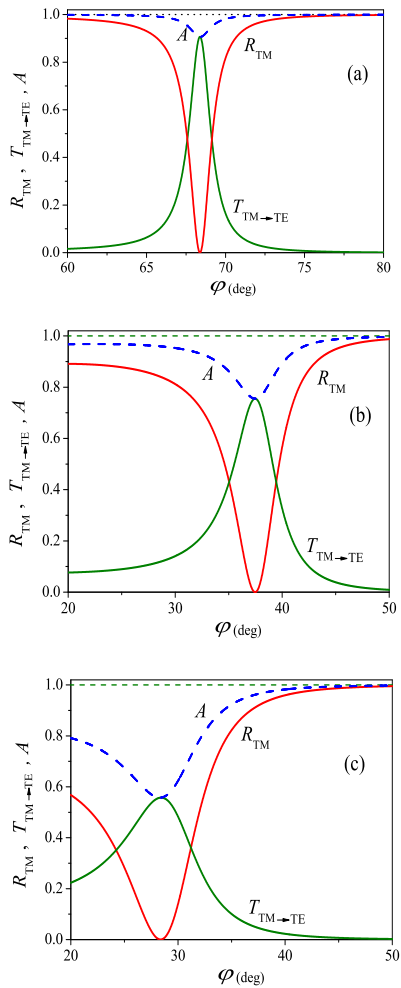}
\caption{\label{Fig6} (Color online) Dependence of the reflectivity $R_{\rm TM}$ (red solid curves), the transformation coefficient $T_{{\rm TM}\rightarrow {\rm TE}}$ (green solid curves), and the absorptivity $A$ (dashed blue curves) on the incident angle $\varphi$ for lossy superconductors for $\Omega = 1$, $\varepsilon_s=16$, $\gamma = 200$, $h=0.5 c/\omega_J$, $\varepsilon_p=11.6$, $\varepsilon_d=1$. The values of other parameters are: (a) $\nu_{ab}=\nu_c =10^{-5}$, $\vartheta = 6.3^\circ$; (b) $\nu_{ab}=\nu_c=10^{-3}$, $\vartheta = 15.8^\circ$; (c) $\nu_{ab}=\nu_c=10^{-2}$, $\vartheta = 26^\circ$.}
\end{figure}

For the lossy superconductor with $\Gamma \neq 0$, the complete resonance suppression of the reflectivity can also be observed. In this case, the energy flux of the incident TM wave is partially reflected as the TE wave. The other part $A$ of the energy [absorptivity $A$ in Eq.~(\ref{A})] is dissipated in the superconductor (see Fig.~\ref{Fig6}).

It is important to note that the phenomenon described here owe the same origin as the wave interaction with an open resonator~\cite{bliokh}. The tunneling of an incident plane wave through an open 1D resonator is characterized by the reflection and transmission coefficients $R$ and $T$ which play the same role as the coefficients $R_{\rm TM}(\varphi, \vartheta)$ and $T_{{\rm TM}\rightarrow {\rm TE}}$ in our analysis. Moreover, the formulas for the coefficients $R$ and $T$ presented in Ref.~\onlinecite{bliokh} are exactly the same as Eqs.~(\ref{RR}) and (\ref{TT}). The role of the leakage parameters $L_{\rm TM}$ and $L_{\rm TE}$ are played by the inverse leakage Q-factors which are related to the transmittances of the two tunnel barriers discussed in Ref.~\onlinecite{bliokh}. However, it is necessary to emphasize a principal novel feature of the phenomenon considered here. We now pay attention to the dual role of the coupling of the electromagnetic fields in the dielectric prism and in the layered superconductor in the conversion of the terahertz wave polarization. First, due to the coupling, the incident TM wave resonantly excites the oblique surface wave. Then, the OSW supplies the TE bulk wave by the energy which passes through the dielectric gap due to the same coupling. Thus, the oblique surface wave serves as an intermediary between the incident TM and the reflected TE waves in the phenomenon of the transformation of the wave polarization.

\section{Conclusion}

We have theoretically studied oblique surface waves which propagate along the interface between a dielectric and a layered superconductor. We consider a geometry in which the interface is perpendicular to the superconducting layers. We have analyzed the case of arbitrary direction of the OSW propagation with respect to the layers. In this general case, the electromagnetic field of the OSWs is a superposition of ordinary and extraordinary waves. We derived the dispersion equation for the OSWs, which is the most general dispersion relation for surface Josephson plasma waves in the geometry considered. We have also shown that the dispersion curves have end-points where the extraordinary mode transforms from evanescent wave to bulk wave propagating deep into the superconductor. In addition, we have analytically solved the problem of the resonance excitation of the OSWs by the attenuated-total-reflection method using an additional dielectric prism. Due  to the strong current anisotropy in the boundary of the superconductor, the excitation of the OSWs is accompanied by an additional important phenomenon. Namely, the electromagnetic field with  orthogonal polarization appears in the wave reflected from the bottom of the prism. We show that, for definite optimal combinations of the problem parameters (the wave frequency, the direction of the incident wave vector, the thickness of the gap between the dielectric prism and superconductor, etc.), there is a complete suppression of the reflected wave with the same polarization as the incident wave. This phenomenon is an analog of the so-called Wood anomalies in the reflectance of isotropic metals known in optics. However, in the isotropic case, the optimal thickness of the dielectric gap between the prism and the metal sample should be chosen for the complete suppression of the reflected wave. However, in the anisotropic case considered here, this phenomenon can be more easily observed by the appropriate choice of the direction of the wave propagation with respect to the superconducting layers. We obtained the conditions for the observation of the complete suppression of the reflectivity and found that, contrary to the isotropic case, this phenomenon can be observed even in the dissipationless limit. In such a regime, a complete conversion of the wave polarization takes place, i.e., 100\% of the  energy flux of the incident TM wave reflects from the bottom of the prism as a wave with TE polarization. Similar conversion phenomena can be observed for the inverse transition, from TE to TM modes, as well as for the transition from incident ordinary waves to extraordinary ones (or viceversa).

\section{Acknowledgements}

We gratefully acknowledge partial support from the ARO, JSPS-RFBR Contract No.~12-02-92100, Grant-in-Aid for Scientific Research (S), MEXT Kakenhi on Quantum Cybernetics, the JSPS-FIRST program, Ukrainian State Program on Nanotechnology, and the Program FPNNN of the NAS of Ukraine (grant No~9/11-H).

\end{document}